\documentclass[aip,amsmath,amssymb,reprint,showkeys]{revtex4-1}

\usepackage{graphicx}
\usepackage{dcolumn}
\usepackage{bm}
\usepackage[utf8]{inputenc}
\usepackage[T1]{fontenc}
\usepackage{mathptmx}
\usepackage{etoolbox}
\usepackage{xcolor}
\usepackage[normalem]{ulem}
\usepackage{comment}

\newcommand{\R}{\mathbf{r}}

\DeclareMathOperator{\erf}{erf}

\makeatletter
\def\@email#1#2{%
 \endgroup
 \patchcmd{\titleblock@produce}
  {\frontmatter@RRAPformat}
  {\frontmatter@RRAPformat{\produce@RRAP{*#1\href{mailto:#2}{#2}}}\frontmatter@RRAPformat}
  {}{}
}%
\makeatother

\begin{document}

\title{Gaussian expansion of Yukawa non-local kinetic energy functionals: application to metal clusters}
\author{Fulvio Sarcinella}
\affiliation{Center for Biomolecular Nanotechnologies, Istituto Italiano di Tecnologia, Via Barsanti 14, 73010 Arnesano (LE), Italy}
\affiliation{Department of Mathematics and Physics “E. De Giorgi”, University of Salento, Via Arnesano, Lecce, Italy}
\author{Szymon \'Smiga}
\affiliation{Institute of Physics, Faculty of Physics, Astronomy and Informatics, Nicolaus Copernicus University in Toru\'n,
ul. Grudzi\c adzka 5, 87-100 Toru\'n, Poland}
\author{Fabio Della Sala}
\affiliation{Institute for Microelectronics and Microsystems (CNR-IMM), Via Monteroni, Campus Unisalento, 73100 Lecce, Italy}
\affiliation{Center for Biomolecular Nanotechnologies, Istituto Italiano di Tecnologia, Via Barsanti 14, 73010 Arnesano (LE), Italy}
\author{Eduardo Fabiano}
\affiliation{Institute for Microelectronics and Microsystems (CNR-IMM), Via Monteroni, Campus Unisalento, 73100 Lecce, Italy}
\affiliation{Center for Biomolecular Nanotechnologies, Istituto Italiano di Tecnologia, Via Barsanti 14, 73010 Arnesano (LE), Italy}
\date{\today}


\begin{abstract}
The development of kinetic energy (KE) functionals is one of the current challenges in  density functional theory (DFT). The Yukawa non-local KE functionals [Phys. Rev. B 103, 155127 (2021)]  have been shown to describe accurately the Lindhard response of the homogeneous electron gas (HEG) directly in the real space, without any step in the reciprocal space. 
However, the Yukawa kernel employs an exponential function which cannot be efficiently represented in conventional Gaussian-based quantum chemistry codes.
Here, we present an expansion of the Yukawa kernel in Gaussian functions. We show that for the HEG this expansion is independent of the electronic density, and that for general finite systems the accuracy can be easily tuned.
Finally, we present results for atomistic sodium clusters of different sizes, showing that simple Yukawa functionals can give superior accuracy as compared to semilocal functionals.
\end{abstract}

\maketitle

\section{Introduction}
Kohn-Sham (KS) Density Functional Theory (DFT) is one of the most used approaches
for the calculation of the electronic properties of quantum systems \cite{dftbookgross,ks,burke2012perspective,becke2014perspective}.
The accuracy of KS-DFT is directly related to the approximations made for the exchange-correlation (XC) functional and  hundreds of different XC functionals have been developed \cite{scuseriaREVIEW05,libxc,headgord2022}.
A linear-scaling alternative to KS-DFT is the  Orbital-Free (OF) DFT\cite{wang2002orbital,ofdft_book,gavini2007quasi}, 
for which different implementations have been made available recently \cite{profess,gpaw14,atlas16,dftpy,golub20,witt_2022}.

In OF-DFT  
the main quantity to be approximated is, instead, the non-interacting kinetic energy (KE) functional
\begin{equation}
T_s = \frac{1}{2} \sum_{i\sigma} \int  |\nabla \phi_{i\sigma}({\bf r})|^2 d^3{\bf r} = T_s[n] \; .
\end{equation}
The KE is known exactly in terms of KS orbitals, which are not available in OF-DFT: thus, one of the biggest challenges in DFT \cite{kara12,witt2018orbital} is the definition of $T_s$ in terms of the electronic density $n$.
Note that the KE functional is also a core quantity in related approaches, such as density-embedding \cite{wesocr,cancio2016visualization} and quantum-hydrodynamic theory \cite{toscano15,ciraci2016quantum,moldabekov2018,bagh21,dellasala22}.
Current approximations to $T_s[n]$ are based on i) semilocal functionals \cite{LC94,perdew2007laplacian,kara09,constantinPRL11,karasiev2013nonempirical,borgooJCTC13,xia2015single,constantin2018semilocal,luo2018simple,letho19,constantin19}
and on ii) non-local functionals with a Lindhard kernel \cite{alonso1978nonlocal,wang1992kinetic,smargiassi1994orbital,garcia1996nonlocal,wang1998orbital,wang1999orbital,zhou2005improving,garcia2008approach,huang2010nonlocal,shin2014enhanced,constantin2018nonlocal,genova2018nonlocal,xu2020,revhc2021}.

Semilocal functionals are explicit functions of real space quantities ($n$,$\nabla n$, $\nabla^2 n$, $\ldots$) and  can be efficiently applied to both  finite and periodic systems. 
Note that finding accurate analytical expressions for semilocal functionals can be very cumbersome: thus, recently, machine-learning techniques have been largely used for this task \cite{goloub19,seino19,meyer20,fuji20,imoto21}.
Despite recent progresses\cite{constantin2018semilocal}, the overall accuracy is quite limited, especially for molecular systems.

Non-local functionals are more accurate but are necessarily defined in the reciprocal space, as no analytical expression nor simple numerical treatment exists for the Lindhard function in the real space \cite{herring1986explicit,wang1992kinetic,kaxi2002,cervera07,ho2008}.
Despite recent advances 
\cite{mi19,xu2020}, calculations of isolated systems
have to be performed in the periodic space with the use of a large
supercell approach (to avoid interactions of periodic replicas).



%
Thus, both classes of functionals have positive and negative features.
Recently, we have introduced a new class of KE functionals, named Yukawa-Generalized Gradient Approximation (yGGA), with the following general form \cite{ygga,ygga_linresp}:
\begin{equation}
T_s^{yGGA} = \int \tau^{TF}(\R) F_s[p(\R),q(\R),y_\alpha(\R)]d^3\R\ ,
\end{equation}
where $\tau^{TF}(\R)=(3/10) n(\R) k_F(\R)^2$ [with $k_F(\R)=(3\pi^2n(\R))^{1/3}$ and $n(\R)$ being the Fermi wave vector and the electron density, respectively] is the Thomas-Fermi (TF) kinetic energy density (KED), $F_s$ is the enhancement factor, $p=|\nabla n|^2/(4k_F^2n^2)$ is the reduced gradient, $q=\nabla^2n/(4k_F^2n)$ is the reduced Laplacian, and
\begin{equation}
y_\alpha(\R) = \frac{3\pi\alpha^2}{4k_F(\R)}u_\alpha(\R) \; \textrm{with} \;u_\alpha(\R)=\int\frac{n(\R')e^{-\alpha k_F(\R)|\R-\R'|}}{|\R-\R'|}d\R' \label{eq:udef}
\end{equation}
is the reduced Yukawa potential (with $\alpha$ being a parameter).
The reduced Yukawa potential is a novel and useful input quantity for the construction of advanced kinetic functionals.
In fact it possesses several useful properties \cite{ygga}: (i) it is a local quantity but it entails non local-features; (ii) it is positive, adimensional, and invariant under the uniform scaling of the density \cite{scaling13}; (iii) it is a good indicator for system-size dependence, in contrast to other semilocal indicators.
The inclusion of a normalized non-local indicator to extend the applicability of Generalized Gradient Approximation (GGA) functionals has been also recently tested for the development of machine-learned XC functionals \cite{cider22} .

A key point of the reduced Yukawa potential $y_\alpha$ is that it yields a non-linear contribution to the linear response function of the HEG, so that the Lindhard function can be well reproduced \cite{ygga,ygga_linresp}.
This is a fundamental improvement with respect to conventional KE functionals based only on semilocal ingredients (such as $p$ and $q$), which yield an incorrect polynomial linear response function \cite{constantin19,ygga,ygga_linresp}. 

Actually, only a few yGGA kinetic functionals have been proposed using the linear ansatz
\begin{equation}
    F_s(p,q,y_\alpha) = \frac{5}{3}p + y_\alpha G(p,q)\ , \label{eq:fdef}
\end{equation}
where $\frac{5}{3}p$ is the von Weizs\"{a}cker (vW) KE enhancement factor. 
Among these we mention the yuk3 and yuk4 functionals \cite{ygga} which are defined by $\alpha=1.3629$ and
\begin{eqnarray}
   G(p,q) = G(x) = T_4(x)\ , &&  \mathrm{yuk3} \label{eq:gfunc} \\
  G(p,q) = T_{3.3}\left(-40p/27\right)T_2\left(40q/27\right)\ ,&&   \mathrm{yuk4}
\end{eqnarray}
where
\begin{eqnarray}
    T_a(x) & = & \frac{4e^{ax}}{a(e^{ax}+1)} + \frac{a-2}{a}\ , \\
    x & = & 40(q-p)/27 \label{eq:xdef}\ .
\end{eqnarray}
The $a$ parameter for the yuk3  ($a=4$) and yuk4 ($a=3.3$ and $2$) functionals have been optimized on jellium clusters\cite{ygga}. 
Current applications of yGGA functionals are limited to spherical systems, where
Eq. (\ref{eq:udef}) can be easily computed. Very recently, applications to extended systems have been presented \cite{witt_2022}.
The calculation of the integral in Eq. \eqref{eq:udef} is not straightforward in quantum chemistry codes, which make use of a Gaussian basis set for the representation of the electronic density: in fact, the integral is not analytical and thus needs to be evaluated numerically, which is computationally expensive.
Instead, the following integral
\begin{equation}
 V(\R)=   \int g_a(\R') g_b(\R') \frac{e^{-a(\R)|\R -\R'|^2}}{|\R -\R'|}  d\R' \ , \label{eq:gint}
\end{equation}
where $g_a$ and $g_b$ are Gaussian basis functions, can be evaluated analytically and thus quite efficiently.
The integral in Eq. (\ref{eq:gint}) is just the (Gaussian) screened electrostatic repulsion of the basis set product $g_a g_b$.
Similar integrals are present in hybrid functionals
with local range separation\cite{kru2008} or in molecular mechanics
with generalized interaction \cite{peels2020}.

The simple substitution in Eq. (\ref{eq:udef}) of the exponential term with a Gaussian one is, however, not a feasible solution because the use of the Gaussian screening in place of the exponential one would alter the linear response properties of the functional, thus making a whole redefinition of the KE functional necessary. For this reason, in this work, we explore a different path and we consider a Gaussian expansion of the original Yukawa kernel in order to preserve the original formulation of the functional and, at the same time, benefit from the computational efficiency of the Gaussian functions.

As a final note, we remark that for the evaluation of the KE potential of yGGA functionals, additional
integrals are required \cite{ygga}. The evaluation of those integrals in a Gaussian basis set requires different routines or
automatic differentiation techniques \cite{witt_2022,kaupp21}.
In this work, we will limit our attention to the evaluation of the KE total energies and kinetic energy density.


\section{Gaussian expansion of the Yukawa kernel}
%

We consider the following Gaussian expansion of the Yukawa kernel:
\begin{equation}\label{eq:gaussian_sum}
\frac{e^{- \omega k_F (\R) |\R - \R'|}}{|\R - \R'|} \simeq \sum_{p = 1}^{M} c_p \frac{e^{- \omega_p k_F^2 (\R) |\R - \R'|^2}}{|\R - \R'|} \ ,
\end{equation}
where $omega$ is a positive parameter (i.e. the analogous of $\alpha$ in Eq. (\ref{eq:udef})), whereas $\omega_p$ and $c_p$ are coefficients to be optimized by minimization of the quantity
\begin{equation}\label{eq:error}
E(\{\omega_p\},\{c_p\};\R) = \displaystyle \int \left( \frac{e^{-\omega k_F (\R) |\R - \R'|}}{|\R - \R'|} - \displaystyle \sum_{p = 1}^{M} c_p \frac{e^{- \omega_p k_F^2 (\R) |\R - \R'|^2}}{|\R - \R'|} \right)^2 \, d\R' \ .
\end{equation}

It is straightforward to show (see Appendix \ref{appa}) that 
\begin{equation}\label{feq}
E(\{\omega_p\},\{c_p\};\R) = \frac{2\pi}{k_F(\R)}F(\{\omega_p\},\{c_p\})\ ,
\end{equation}
where
%
%
\begin{equation}
F(\{\omega_p\},\{c_p\})  =  \frac{1}{\omega} 
+ \sqrt{\pi}  \sum_{p ,q= 1}^{M} c_p A_{pq} c_q 
-2\sqrt{\pi}\sum_{p = 1}^{M} c_p b_p\ ,
\end{equation}
with
\begin{eqnarray}
A_{p q} & = & \frac{1}{\sqrt{\omega_p + \omega_{q}}}\ , \\
b_{p} & = & e^{\omega^2/(4 \omega_{p})} \frac{1}{\sqrt{\omega_{p}}} \left[ 1 - \erf\left( \frac{\omega}{2 \sqrt{\omega_{p}}} \right) \right] \ .
\end{eqnarray}

Therefore, we can neglect the shape factor $2\pi/k_F$ and focus on the factor $F(\{\omega_p\},\{c_p\})$. Because the latter is independent of the position and  the density, the minimization will result in a universal approximation of the Yukawa kernel (within the chosen Gaussian expansion space).
Note that such an optimization for the kernel is equivalent to 
optimizing $y_\alpha(\R)$ for a uniform electronic density, i.e., for the HEG.

To minimize $F(\{\omega_p\},\{c_p\})$, we consider in the first step its variation with respect to the coefficients $c_p$, setting it equal to zero.
%
%
%
Then, we readily obtain the equation 
\begin{equation}\label{eq:matrix}
\sum_{q=1}^{} A_{p q} c_q = b_{p} \ .
\end{equation}
The $F$ function at the optimized $\{c_p\}$ coefficients is thus
\begin{equation}
\bar{F}(\{\omega_p\})  = 
\frac{1}{\omega} 
- \sqrt{\pi}  \sum_{p ,q= 1}^{M} b_p A^{-1}_{pq} b_q \ ,
\end{equation}
i.e. a non-linear function of $\{ \omega_p \}$ only.
%


To minimize $\bar{F}(\{\omega_p)\}$ we employ the following strategy. We define a set of $M$ $\omega_p$ parameters that form a geometric series \cite{gtobas} within the interval [$\omega_{min}$,$\omega_{max}$]. Using this set the value of $\bar{F}$ depends only on two parameters, $\omega_{min}$ and $\omega_{max}$. We then minimize $\bar{F}$ by scanning over a wide range of $\omega_{min}$ and $\omega_{max}$ values. After this accomplishment we have a quasi-optimal set of $\omega_p$ (distributed in a geometric series between the optimized $\omega_{min}$ and $\omega_{max}$). In a final step we use this quasi-optimal set as a starting point for a further multivariate minimization of $\bar{F}(\omega_p)$ where now all the $\omega_p$ parameters are free to vary.

\begin{figure}
    \centering
    \includegraphics[width=0.45\textwidth]{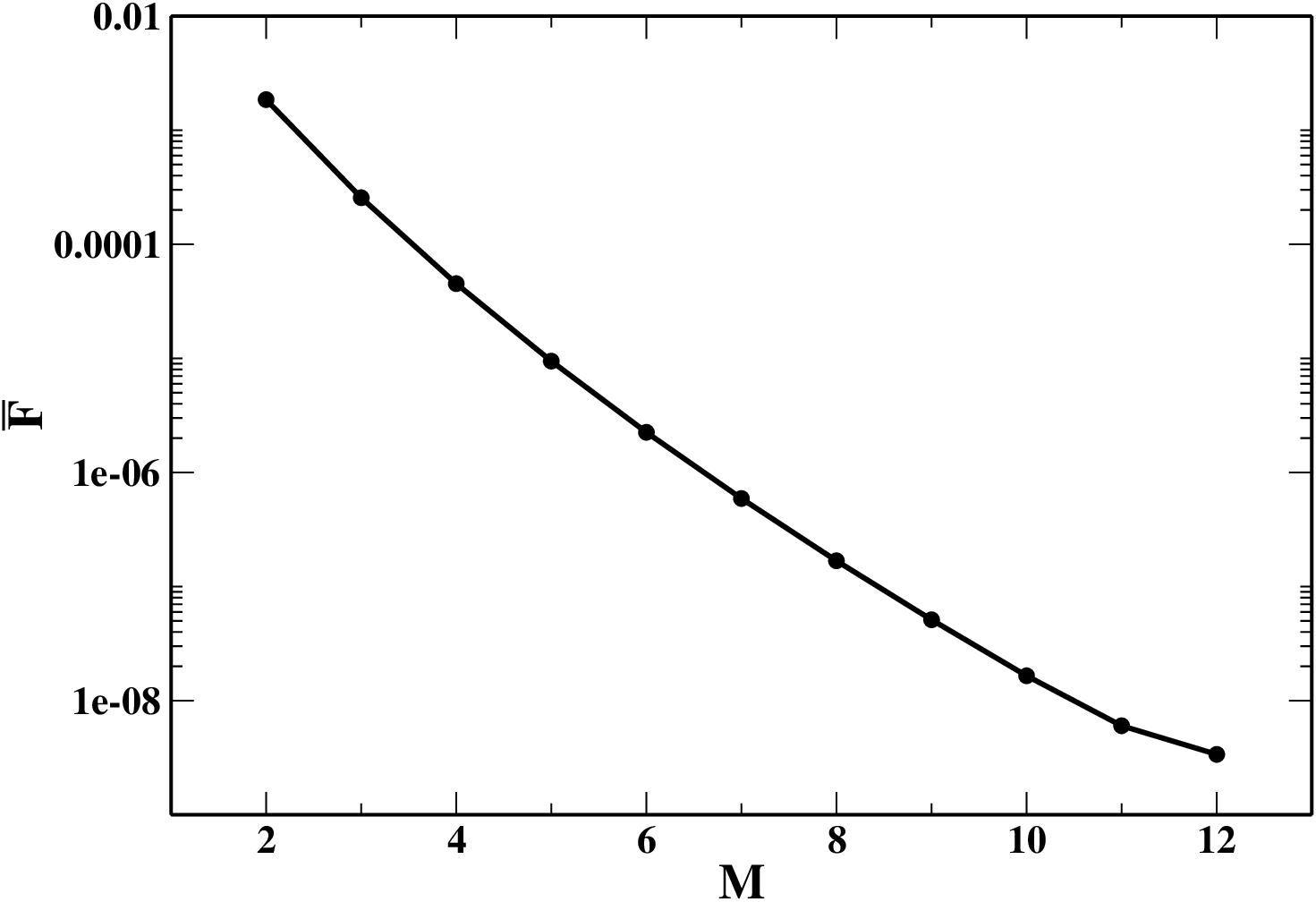}
    \caption{Values of $\bar{F}$ for different numbers $M$ of Gaussians.}
    \label{fig_N}
\end{figure}

The results of the minimization of $\bar{F}$ for different values of $M$ are reported in Fig. \ref{fig_N}. 
Note that in this study we only consider the fixed value $\alpha = 1.3629$ as in the yuk3 and yuk4 functionals; other values of the $\alpha$ parameter could be easily considered in a similar manner but they are not investigated in this work.
The plot shows that the accuracy of the approximation increases quite fast (exponentially) with the number of Gaussians, whereas its cost scales only linearly.
However, for relatively large values of $M$ ($M\ge 10$), numerical errors
occur in the solution of the linear system in Eq. (\ref{eq:matrix}), as the matrix $A$ has a very large condition number, and the improvement is only marginal.

Thus, from a pragmatic point of view we can select three levels of approximation with increasing computational cost: 
loose ($\bar{F}\approx 10^{-4}$, $M=3$), 
medium ($\bar{F}\approx 10^{-6}$, $M=6$), 
and high ($\bar{F}\approx 10^{-8}$, $M=9$). The corresponding values of $\omega_p$ and $c_p$ are reported in Tab. \ref{tab_omega}.
\begin{table}
    \centering
    \begin{tabular}{rrrrrr}
    \hline\hline
    \multicolumn{2}{c}{$M=3$} & \multicolumn{2}{c}{$M=6$} & \multicolumn{2}{c}{$M=9$}\\
    \cline{1-2}\cline{3-4}\cline{5-6}
        \multicolumn{1}{c}{$\omega_p$} & \multicolumn{1}{c}{$c_p$} & \multicolumn{1}{c}{$\omega_p$} & \multicolumn{1}{c}{$c_p$} & \multicolumn{1}{c}{$\omega_p$} & \multicolumn{1}{c}{$c_p$} \\
    \hline
 0.3450 & 0.27663 & 0.1891 & 0.08688 & 0.1369 & 0.03314 \\
 2.0803 & 0.43380 & 0.6077 & 0.27877 & 0.3450 & 0.16366 \\ 
25.1512 & 0.24289 & 2.2002 & 0.28762 & 0.9311 & 0.24504 \\
        &        & 9.6803 & 0.18982 & 2.6728 & 0.21743 \\
        &        & 58.6704 & 0.10168 & 8.4791 & 0.15181 \\
        &        & 712.5598 & 0.04648 & 30.7659 & 0.09372 \\
        &        & & & 135.5610 & 0.05306 \\
        &        & & & 822.0016 & 0.02737 \\
        &        & & & 9984.8049 & 0.01242 \\
 \hline\hline
    \end{tabular}
    \caption{Optimized values of the $\omega_p$ and $c_p$ parameters for different values of $M$ (i.e. the number of Gaussians).}
    \label{tab_omega}
\end{table}

To benchmark the effectiveness of the approximations we  consider the values of several indicators computed for the three model spherical one-electron densities\cite{const2011}
\begin{equation}\label{e_model_dens}
n_H(r) = \frac{e^{-2r}}{\pi}\; ,\; n_G(r)=\frac{e^{-r^2}}{\sqrt{\pi}^3}\; ,\; n_C(r) = \frac{(1+r)e^{-r}}{32\pi}\ ,
\end{equation}
which are models for atomic, molecular and solid-state densities.
The indicators are designed to assess the effect of the approximation $\Delta y_\alpha(\R) = y_\alpha^G(\R) - y_\alpha(\R)$, where $y_\alpha^G(\R)$ denotes the quantity $y_\alpha$ computed using the Gaussian approximation for the Yukawa potential.
For any linear yGGA functional with general form
\begin{equation}
T_s^\mathrm{yGGA} = T_{vW} + \int\tau^{TF}[n](\R)y_\alpha[n](\R)G(p,q)d\R\ , 
\end{equation}
where $T_{vW} = 5p/3$, the error induced by $\Delta y_\alpha$ is
\begin{equation}
\Delta T_s^\mathrm{yGGA} = \int\tau^{TF}(\R)G(p,q)\Delta y_\alpha(\R)d\R\ .
\end{equation}
Hence, we consider the two indicators corresponding to $G(p,q)=1$ and $G(p,q)=G^\mathrm{yuk3}(p,q)$. That is
\begin{eqnarray}
\label{e_epsilon}
\epsilon & \equiv & \int \tau^{TF}(\R)\Delta y_\alpha(\R)d\R\ , \\
\label{e_zeta}
\zeta & \equiv & \int \tau^{TF}(\R)G^\mathrm{yuk3}(p,q)\Delta y_\alpha(\R)d\R\ .
\end{eqnarray}
The indicator $\epsilon$ is not only a measure of the error for the simplest linear yGGA, but provides also an indication of the density weighted error on $\Delta y_\alpha$.

\begin{figure}
    \centering
    \includegraphics[width=0.95\columnwidth]{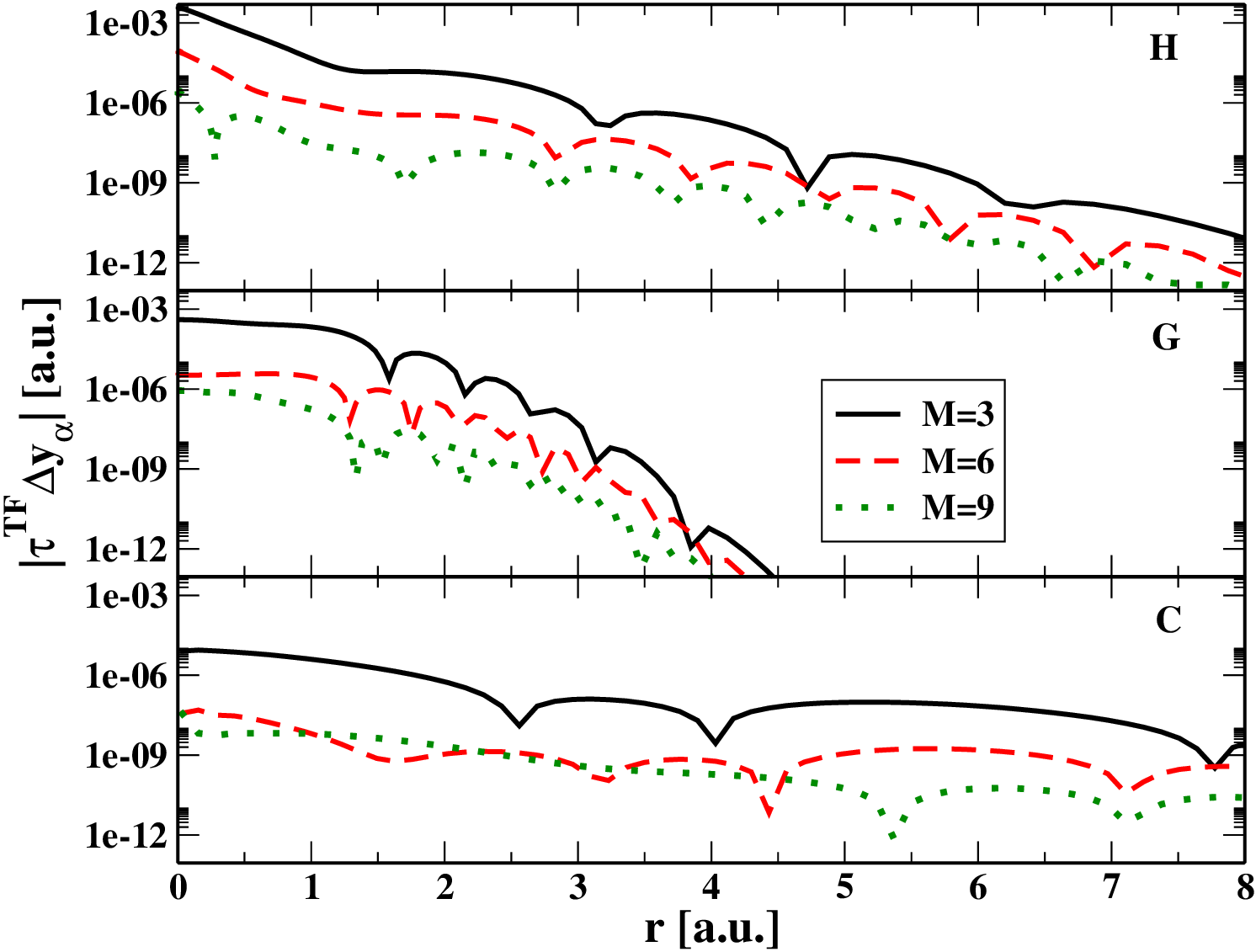}
    \caption{Integrand of the indicator $\epsilon$ [Eq. (\ref{e_epsilon})] for the three model densities H, G, and C.}
    \label{fig_epsilon_error}
\end{figure}

The values of the indicators for the three model densities at the various levels of approximation are reported in Tab. \ref{tab_model_dens} and the space profile of the corresponding integrands are shown in Fig. \ref{fig_epsilon_error}.
\begin{table}
    \centering
    \begin{tabular}{llrrr}
    \hline\hline
    Density & Indicator & \multicolumn{3}{c}{$M$} \\
    \cline{3-5}
    & & 3 & 6 & 9 \\
    \hline
H & $\epsilon$ & -1.851E-3 & -3.199E-5 & 2.454E-7 \\
  & $\zeta$    & -9.314E-4 & -1.608E-5 & 1.267E-7 \\
G & $\epsilon$ &  1.899E-3 & -1.572E-5 & 2.133E-6 \\
  & $\zeta$    &  9.690E-4 &  -8.103E-6 & 1.099E-6 \\
C & $\epsilon$ & -1.308E-4 & -9.477E-7 & 2.721E-7 \\ 
  & $\zeta$    & -6.589E-5 & -4.748E-7 & 1.377E-7 \\
  \hline\hline
    \end{tabular}
    \caption{Values of the indicators $\epsilon$ [Eq. (\ref{e_epsilon})] and $\zeta$ [Eq. (\ref{e_zeta})] for the three model densities of Eq. (\ref{e_model_dens}).}
    \label{tab_model_dens}
\end{table}
These numbers confirm that, at different density regimes, the expected errors are quite small already with the lightest approximation ($M=3)$ and become very small for larger values of $M$.

\section{Computational details and implementation}
To test the Gaussian approximation of the Yukawa potential, we have implemented it into the in-house code \verb|jkinplot|, which is able to handle systems with radial symmetry. Thus, we could compute both the exact and the approximate Yukawa contributions for several jellium spheres and the sodium atom (see appendix \ref{appb} for details) as well as various kinetic approximations.
We used the same setup as in Ref. \onlinecite{ygga} for these calculations.

To test more realistic systems, i.e., sodium clusters, we implemented the Gaussian approximation of the Yukawa potential into the locally modified version of  ACESII\cite{acesII} quantum chemistry code. The integrals have been realized according to Ref. \onlinecite{peels2020} (see Sec. F1 in Supporting information file) and implemented in the plot module program.

The sodium clusters geometries were taken from Ref.~\onlinecite{na_clusters} and reoptimized using the def2-TZVP basis set and the Local Density Approximation (LDA) exchange-correlation functional.
All calculations employed the LANL08 basis set with the corresponding effective core potential (ECP) \cite{lanl08} so we finally have 1 electron per Na atom. A simple cubic Cartesian grid, enclosing the cluster, such that the electron density on all the cube facets is below the threshold $10^{-6}$ a.u., has been employed. We chose a grid step of 0.5 bohr, which is sufficient to grant converged kinetic energies up to 1e-3 Ha, because of the absence of core electrons and the metallic character of the sodium clusters considered. 

The yGGA functionals are compared to different local and semilocal kinetic functionals, such as the LDA TF functional, the functionals employing the full vW term, i.e.\ TFvW, PGS~\cite{constantin2018nonlocal}, PG1~\cite{constantin2018nonlocal}, VT84f~\cite{karasiev2013nonempirical}, the gradient expansions GE2 and GE4, and one functional based on the asymptotic expansion of the semiclassical neutral atom, revAPBEk~\cite{constantinPRL11}.

\section{Results}
In this section we consider several results where the Gaussian approximation of the Yukawa potential has been used to generate the $y_\alpha$ ingredient and compute yuk3 and yuk4 kinetic energies.
Initially, we consider jellium spheres, where the Yukawa potential can also be computed exactly, due to the spherical symmetry. Thus, we can accurately benchmark our approximation.
Then, we consider a set of sodium clusters that can only be simulated using the Gaussian approximation proposed in this paper.


\subsection{Jellium spheres}
Table \ref{tab:ene_jellium} reports the kinetic energy computed with the yuk3 kinetic functional for various jellium spheres together with the errors on this quantity obtained employing the Gaussian approximation with $M=3,6,9$ Gaussians respectively.
\begin{table}
\begin{center}
\caption{\label{tab:ene_jellium} Kinetic energy (Ha) for jellium clusters of different sizes ($N$ = $40$, $92$, $138$, $254$, $438$) and Wigner-Seitz radii ($r_s$ = $2$, $3$, $4$, $5$, $6$), obtained with KS calculation ($E_\textnormal{KS}$) and according to the kinetic functional yuk3 ($E_\textnormal{yuk3}$). The columns on the right contain the errors between $E_\textnormal{yuk3}$ and the kinetic energies computed with yuk3, but employing the Gaussian expansion, for three different numbers of Gaussian function ($M=3, 6, 9$). 
The last lines report the Mean Absolute Error (MAE) and Mean Absolute Relative Error (MARE) for yuk3 with respect to KS as well as the MAE and MARE of the various approximations with respect to the "exact" yuk3 ($E_\textnormal{yuk3}$).}
\begin{ruledtabular}
\begin{tabular}{rrr|rrr}
$N$, $r_s$ & $E_\textnormal{KS}$ & $E_\textnormal{yuk3}$ & M=3 & M=6 & M=9\\
\toprule
40, 2 & 8.834 & 8.705 & 0.246 & 0.018 & 0.002 \\
40, 3 & 4.255 & 4.201 & 0.114 & 0.008 & 0.001 \\
40, 4 & 2.529 & 2.502 & 0.065 & 0.005 & 0.001 \\
40, 5 & 1.690 & 1.676 & 0.042 & 0.003 & 0.000 \\
40, 6 & 1.217 & 1.211 & 0.030 & 0.002 & 0.000 \\
92, 2 & 21.979 & 21.578 & 0.739 & 0.065 & 0.009 \\
92, 3 & 10.282 & 10.152 & 0.334 & 0.029 & 0.004 \\
92, 4 & 5.990 & 5.943 & 0.190 & 0.017 & 0.002 \\
92, 5 & 3.941 & 3.928 & 0.122 & 0.011 & 0.001 \\
92, 6 & 2.802 & 2.804 & 0.085 & 0.007 & 0.001 \\
138, 2 & 33.420 & 32.878 & 1.221 & 0.113 & 0.016 \\
138, 3 & 15.545 & 15.331 & 0.549 & 0.051 & 0.007 \\
138, 4 & 9.025 & 8.926 & 0.311 & 0.029 & 0.004 \\
138, 5 & 5.924 & 5.875 & 0.200 & 0.018 & 0.003 \\
138, 6 & 4.204 & 4.181 & 0.139 & 0.013 & 0.002 \\
254, 2 & 63.491 & 62.429 & 2.513 & 0.246 & 0.036 \\
254, 3 & 29.214 & 28.797 & 1.124 & 0.110 & 0.016 \\
254, 4 & 16.839 & 16.642 & 0.634 & 0.062 & 0.009 \\
254, 5 & 10.990 & 10.890 & 0.406 & 0.040 & 0.006 \\
254, 6 & 7.762 & 7.711 & 0.282 & 0.028 & 0.004 \\
438, 2 & 110.857 & 109.405 & 4.678 & 0.474 & 0.072 \\
438, 3 & 50.773 & 50.112 & 2.086 & 0.211 & 0.032 \\
438, 4 & 29.175 & 28.825 & 1.175 & 0.119 & 0.018 \\
438, 5 & 18.994 & 18.794 & 0.752 & 0.076 & 0.011 \\
438, 6 & 13.387 & 13.267 & 0.523 & 0.053 & 0.008 \\
\hline
 MAE & & 0.254 & 0.742 & 0.072 & 0.011  \\
MARE (\%) & & 1.06 & 3.45 & 0.32 & 0.04\\
\end{tabular}
\end{ruledtabular}
\end{center}
\end{table}
Inspecting the data, we see that the errors induced by the Gaussian approximation are quite significant when only $M=3$ Gaussians are employed, but they immediately drop to much smaller values for $M=6$ and especially $M=9$.
In fact, the mean absolute error (MAE) due to the introduction of the Gaussian approximation is, for $M=6$, about four time smaller than the intrinsic MAE of the yuk3 functional (i.e. the mean difference $|E_\textnormal{yuk3}-E_\textnormal{KS}|$); when $M=9$ Gaussians are used, the error of the approximation is one order of magnitude smaller than the intrinsic one. 
Similar considerations apply also for the mean absolute relative error (MARE).

The performance of the Gaussian approximation can be further analysed by considering its behavior for the individual systems. In this case we find that the inaccuracies grow slightly with the number of electrons as well as for smaller values of the Wigner-Seitz parameter $r_s$ (i.e., for larger densities). The increase is, however, quite limited such that even the worst-behaving system, the jellium sphere with 438 electrons and $r_s=2$, displays an error of only 0.07 (0.47) Ha for the approximation with $M=9$ ($M=6$) Gaussians. This must be compared with the intrinsic accuracy of the yuk3 for this case, that is 1.45 Ha.

\begin{figure*}[hbt]
    \centering
    \includegraphics[width=0.9\textwidth]{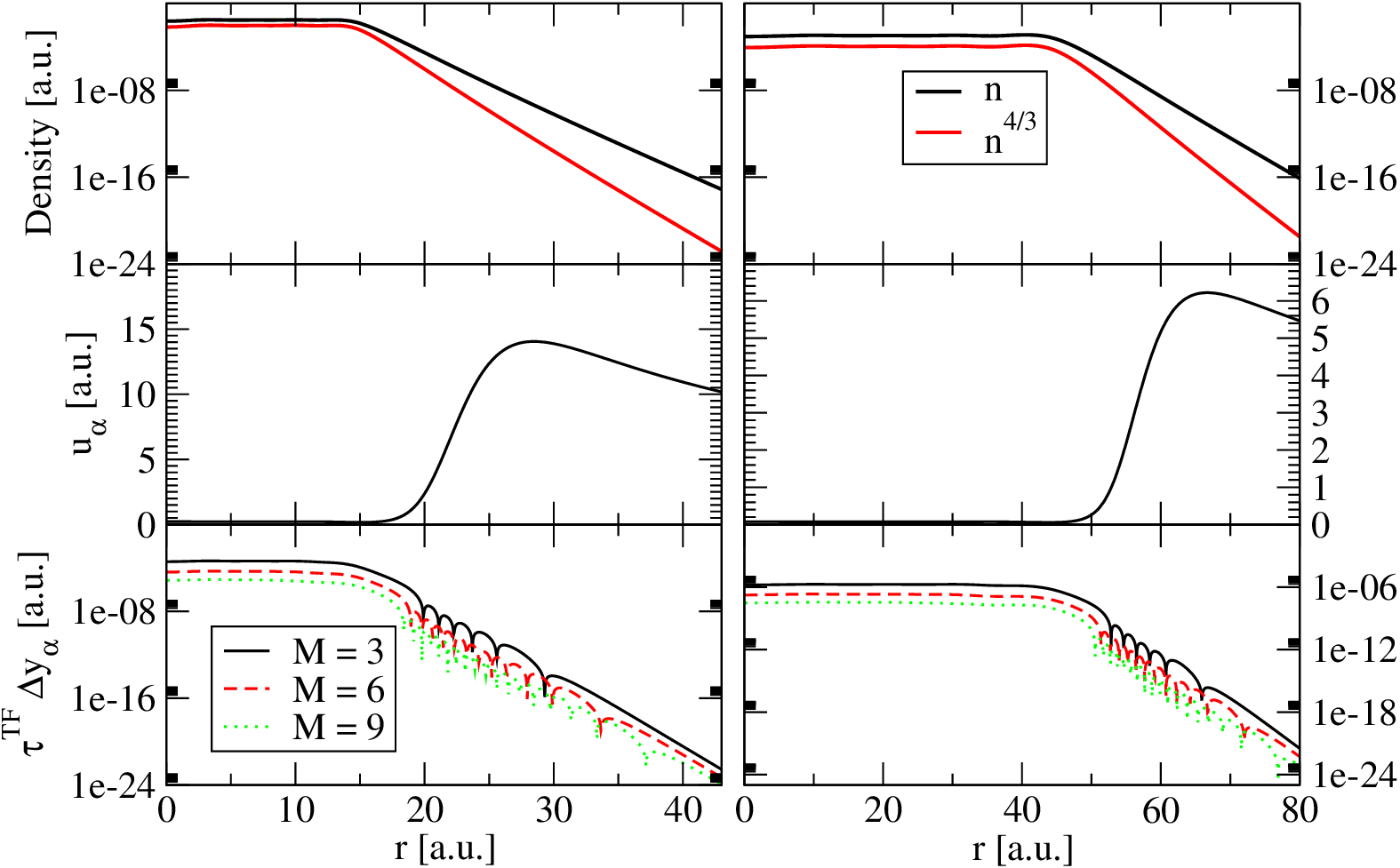}
    \caption{Density (upper panels), $u_\alpha$ (middle panels) and integrand of the indicator $\epsilon$ [Eq. (\ref{e_epsilon})] (lower panels) for three numbers of Gaussian ($M=3, 6, 9$) for the jellium cluster with 438 electrons and $r_s=2$ (left) and $r_s=6$ (right).} 
    \label{fig_jellium}
\end{figure*}

The fact that the errors grow slightly for systems with larger densities may seem counter-intuitive with what one could expect from Eq. (\ref{feq}), which includes a shape factor $2\pi/k_F$ (i.e., the error on the bare Yukawa approximation grows for \emph{smaller} densities). However, we need to recall [see Eqs. (\ref{e_epsilon}) and (\ref{e_zeta})] that the error induced in the kinetic functional is of the order of
$\tau^{TF}y_\alpha$, thus it changes as $\tau^{TF}/k_F\approx n^{4/3}$. This behavior is confirmed by the data plotted in Fig. \ref{fig_jellium}.
%
For this reason the regions with small density tend to contribute less to the functionals inaccuracies and overall the errors compensate such that, in fact, they finally grow linearly with the number of electrons, but with a very small prefactor (about 2e-4 for $M=9$ and about 1e-3 for $M=6$).

\subsection{Sodium clusters}
The results for jellium clusters, reported in the previous subsection, indicate that the Gaussian approximation of the Yukawa potential may be sufficiently accurate to allow yGGA calculations in diverse systems.
Thus, we are now in the position of being able to test yGGA kinetic functionals on atomistic systems beyond spherical symmetry. This is what we attempt in this section, where we employ this approximation to compute the kinetic energies of various sodium clusters.
However, the currently available yGGA functionals have not been developed to treat the density cusp present at the core of the atoms.
Thus, we will focus on the valence electrons only using ECPs in our calculations. In particular, for the sodium atom we will use just one valence electron.

\begin{figure}[hbt]
    \centering
    \includegraphics[width=0.45\textwidth]{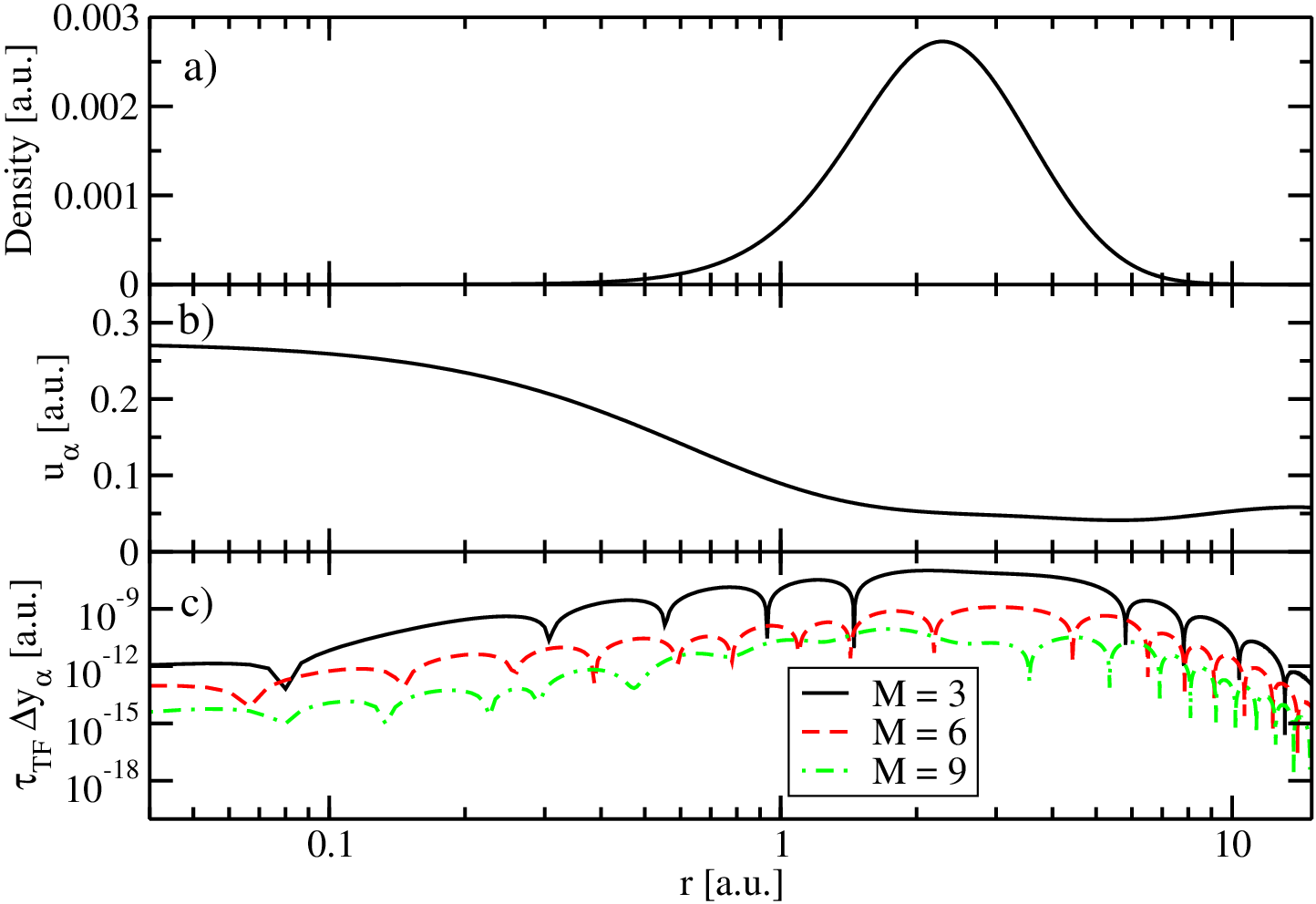}
    \caption{Sodium atom with ECP: a) density, b) the screened Yukawa potential ($u_\alpha$, with $\alpha=1.3629$) and c) integrand of the indicator $\epsilon$ for three numbers of Gaussians ($M = 3, 6, 9$).}
    \label{fig_atom}
\end{figure}

As a preliminary test, we thus consider a single sodium atom, and in  Fig. \ref{fig_atom},  we report the density, the screened Yukawa potential (used in the yuk3 functional)
and errors due to the Gaussian expansions.

%
The plot shows that, although the density shape is rather different from the jellium one considered so far, the error induced by the Gaussian approximation is very small, especially when $M=9$ Gaussians are used to expand the Yukawa kernel.

\begin{figure}[t]
    \centering
    \includegraphics[width=0.45\textwidth]{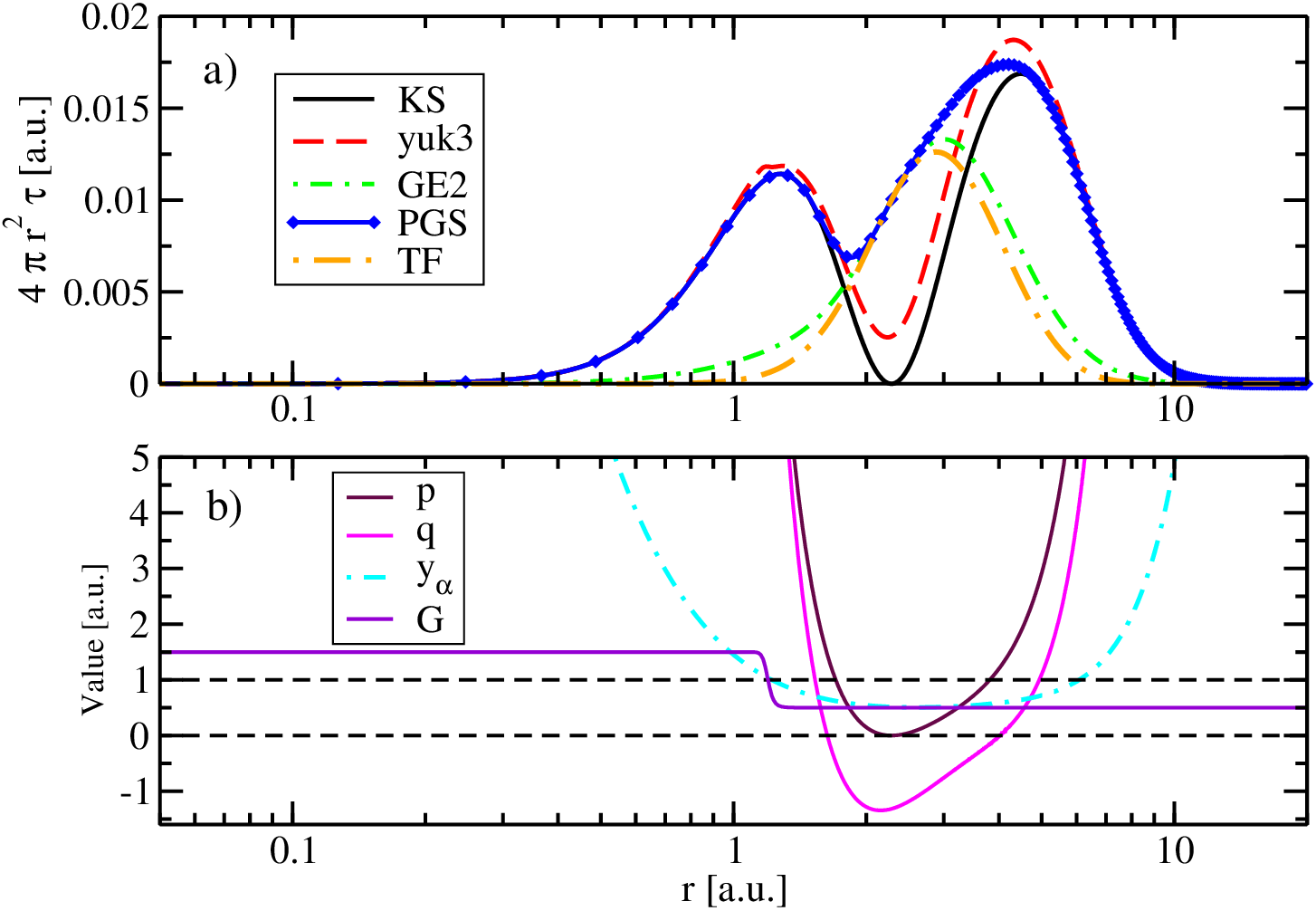}
    \caption{Sodium atom with ECP: a) spherical averaged kinetic energy density for different functionals, b) values of the indicators $p$, $q$, $y_\alpha$ and the function G, see Eq. \ref{eq:gfunc}. In the core and in the tail the density vanishes, thus $p$, $q$, $y_\alpha$ diverge.}
    \label{fig_tau_atom}
\end{figure}

Here it is also worth to note that the yuk3 functional is quite accurate in reproducing the KE of the sodium atom, as shown
in Fig. \ref{fig_tau_atom}, where we report the kinetic energy density for different functionals. In this case (just one electron) the exact KS corresponds to the vW functional. Thus, functionals without the full vW term (i.e. TF and GE2) are quite inaccurate.
In particular,
large differences among functionals are related to the description of the density peak at $r=2.7$ a.u.,%
 where $p$ is vanishing and $q$ is negative.
GE2 and PGS simply recover TF at this point, as  $p=0$ at the peak.
Instead yuk3 gives a very small value, as both G and $y_\alpha$ are less than 1, see Fig. \ref{fig_tau_atom}b. In particular, G is less than 1 because $q$ is negative, and thus  $x$ [see Eq. \eqref{eq:xdef}] is negative, whereas    $y_\alpha$ is less than 1 as it includes a system-size dependence \cite{ygga}.
Thus the term  $y_\alpha G$ in the yuk3 functional, see Eqs. \eqref{eq:fdef} and \eqref{eq:gfunc}, is very small for this one electron system.

Then, we report in Table \ref{tab:ene_cluster} the kinetic energy errors of various functionals, ranging from LDA to yGGA, for several sodium clusters.
\begin{table}[b]
\begin{center}
\caption{\label{tab:ene_cluster} Absolute values of the relative errors, in percent, for the kinetic energies, according to different kinetic functionals, for sodium clusters with different number of atoms $N$. The last column contains the average value for each cluster. The best result for each column is highlighted in bold.}
\begin{ruledtabular}
\begin{tabular}{lccccccc}
$N$ & 16 & 20 & 24 & 30 & 34 & 40 & Avg\\
\toprule
TF & 23.88 & 21.99 & 21.90 & 21.12 & 20.63 & 20.18 & 21.62 \\
TFvW & 27.51 & 24.95 & 23.78 & 22.46 & 21.78 & 20.40 & 23.48 \\
GE2 & 18.17 & 16.78 & 16.83 & 16.28 & 15.92 & 15.68 & 16.61 \\
GE4 & 25.82 & 19.49 & 23.39 & 23.09 & 19.89 & 19.08 & 21.79 \\
yuk1 & 12.89 & 15.67 & 16.72 & 18.11 & 18.84 & 20.13 & 17.06 \\
yuk3 & \bf{4.50} & \bf{6.18} & \bf{7.12} & 7.80 & 8.16 & 8.62 & \bf{7.06} \\
yuk4 & 7.91 & 8.90 & 9.66 & 10.19 & 10.38 & 10.84 & 9.65 \\
PGS & 9.55 & 8.84 & 7.88 & \bf{7.28} & \bf{6.94} & \bf{6.26} & 7.79 \\
PG1 & 13.35 & 12.24 & 11.27 & 10.53 & 10.14 & 9.31 & 11.14 \\
VT84f & 20.90 & 19.04 & 17.93 & 16.83 & 16.26 & 15.06 & 17.67 \\
revAPBEk & 19.10 & 17.68 & 17.68 & 17.10 & 16.72 & 16.45 & 17.45 \\
\end{tabular}
\end{ruledtabular}
\end{center}
\end{table}
The data show that yGGAs, especially yuk3, are quite accurate for sodium clusters, being competitive and slightly better than the best meta-GGAs, twice as better than most GGAs (e.g. GE2) and more than three times better than GE4 and TF.
These are quite encouraging results for further development of the yGGA functionals. Note that the yuk3 has no empirical parameter fitted on atomic  systems.

\begin{figure}
    \centering
    \includegraphics[width=0.45\textwidth]{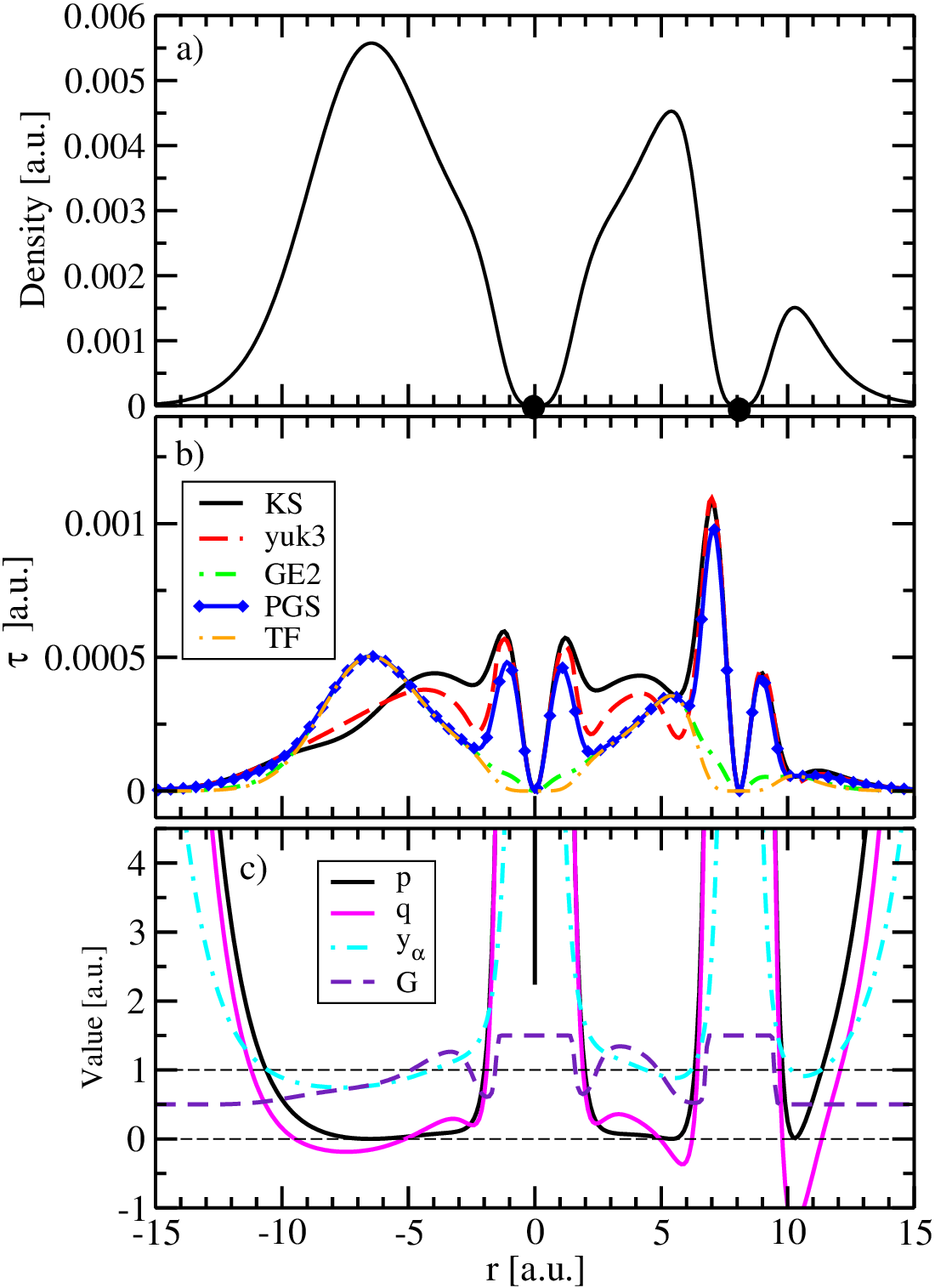}
    \caption{Properties of a Sodium cluster with 16 atoms along a line joining two atoms (indicated by black circles): a) electronic density, b) kinetic energy density $\tau$ for different functionals, c) values of the indicators $p$, $q$, $y_\alpha$ and the function G, see Eq. \ref{eq:gfunc}. 
    }
    \label{fig_tau_cluster}
\end{figure}

The good performance of yuk3 can be traced back to its superior ability to describe the valence region of the sodium atoms. This is illustrated in Fig. \ref{fig_tau_cluster}b) where the yuk3 kinetic energy density is compared to the exact KS one and to other conventional functionals.
All functionals reduce to almost the same value at the atomic core, where
the density and the gradient are vanishing (due to the ECPs).
Near the core, TF and GE2 fail to reproduce the first large oscillation, which is instead reproduced by PGS and yuk3, as also shown in Fig. \ref{fig_tau_atom}.
Another important region is the main density peak at about $r=-6.5$ a.u..
Here $p=0$ and thus TF, GE2, PGS all give the same $\tau$, which is however much larger than the exact KS one, which is instead well reproduced by the yuk3 functional.
At $r=-6.5$ a.u. we have that both $y_\alpha$ and the functional G are less than zero (see Fig. \ref{fig_tau_atom}c), thus the correct KED, smaller than the TF one, is obtained. 
Overall, the yuk3 curve nicely follows all the peaks of the exact KS: this is not the case for PGS, which nevertheless gives accurate (due to error balancing) total energies.

Finally, we recall that the KED is not uniquely defined: for example a term linear in $q$ can be added, without changing the total energy. Despite better KED can be obtained adding a Laplacian term\cite{yang86,garcia07,smiga17}, the overall KED might not be always positive. Thus, in Figs. \ref{fig_atom} and \ref{fig_tau_cluster} we have compared the positive definite KED, which is well defined.

%
%

%

\section{Conclusions}
In this work we have shown how to perform KE calculations with Yukawa based functionals in conventional Gaussian-based quantum chemistry codes.
We show that the Yukawa kernel can be expanded in Gaussian functions, with  universal (i.e. independent of the density) coefficients and exponents.
With M=9 Gaussians the Yukawa potential can be reproduced with negligible errors as compared to reference calculations, for different systems (one-electron density, jellium clusters, sodium atom).

We then tested the yuk3 functional on sodium clusters,  using pseudopotentials, as the yuk3 functional cannot reproduce the electronic cusp correctly.
The results for the yuk3 functional show a non-trivial high accuracy for total kinetic energy.
We found that the non-locality and the size-extensivity of the $y_\alpha$ ingredient plays a key role in this context.
Considering that the yuk3 functional does not include any empirical parameter fitted on atomic systems and it represents only the simplest form (a linear one) of yGGA functional, a significant improvement can be expected for more sophisticated yGGA functionals.

\begin{acknowledgments}
CNR-IMM acknowledges the financial support from ICSC -- Centro
Nazionale di Ricerca in High Performance Computing, Big
Data and Quantum Computing, funded by European Union
-- NextGenerationEU -- PNRR. 
S.\'S. thanks the Polish National Science Center for the partial financial support under Grant No. 2020/37/B/ST4/02713.
\end{acknowledgments}

\bibliography{main}

\appendix

\section{Derivation of Eq. (\ref{feq})}
\label{appa}
We apply to Eq. (\ref{eq:error}) the variable substitution $\R - \R' = \mathbf{x}$ to obtain
\begin{equation}
E = \int \left( \frac{e^{-\omega k_F (\R) |\mathbf{x}|}}{|\mathbf{x}|} - \displaystyle \sum_{p = 1}^{M} c_p \frac{e^{- \omega_p k_F^2 (\R) |\mathbf{x}|^2}}{|\mathbf{x}|} \right)^2 \, d\mathbf{x}
\end{equation}
Hence,
\begin{multline}
E = 4 \pi \int_0^\infty \left(e^{-\omega k_F (\R) x} - \displaystyle \sum_{p = 1}^{M} c_p e^{- \omega_p k_F^2 (\R) x^2} \right)^2 \, dx \\
= 4 \pi \left[\vphantom{\int}\right. \underbrace{\int_0^\infty e^{-2 \omega k_F(\R) x} \, dx}_{\text{I}} \\
+ \underbrace{\int_0^\infty \left( \sum_{p = 1}^{M} c_p e^{- \omega_p k_F^2 (\R) x^2} \right)^2 \, dx}_{\text{II}}\\
\underbrace{- 2 \sum_{p = 1}^{M} c_p \int_0^\infty e^{- \omega_p k_F^2 (\R) x^2 - \omega k_F(\R) x}}_{\text{III}} \left.\vphantom{\int}\right] \ .
\end{multline}
We now solve the three terms I, II and III:
\begin{equation}\label{eq:I}
\text{I} = \int_0^\infty e^{-2 \omega k_F(\R) x} \, dx = \frac{1}{2 \omega k_F (\R)} \ ,
\end{equation}
\begin{multline}\label{eq:II}
\text{II} = \int_0^\infty \left( \sum_{p = 1}^{M} c_p e^{- \omega_p k_F^2 (\R) x^2} \right)^2 \, dx \\
= \sum_p^M \sum_q^M \int_0^\infty c_p c_q e^{- (\omega_p + \omega_q) k_F^2(\R) x^2} \, dx \\
= \frac{\sqrt{\pi}}{k_F(\R)} \sum_p^M \sum_q^M \frac{c_p c_q}{\sqrt{\omega_p + \omega_q}} \ ,
\end{multline}
\begin{multline}\label{eq:III}
\text{III} = - 2 \sum_{p = 1}^{M} c_p \int_0^\infty e^{- \omega_p k_F^2 (\R) x^2 - \omega k_F(\R) x} \\
= -\frac{\sqrt{\pi}}{k_F(\R)} \sum_{p = 1}^{M} c_p \frac{e^{\omega^2/(4 \omega_p)}}{\sqrt{\omega_p}} \left[ 1 - \erf\left(\frac{\omega}{2 \sqrt{\omega_p}}\right) \right] \ .
\end{multline}
Putting all the terms together we finally obtain Eq. (\ref{feq}).

\vspace*{1ex}

\section{Gaussian screened Coulomb integrals in spherical symmetry}
\label{appb}
The integrals containing the Yukawa kernel
can be easily computed numerically for spherical systems.
For conventional integrals with the exponential screening, see Appendix B of Ref. \onlinecite{ygga}.
%
For integrals in which the Gaussian kernel appears,
the computation for systems with radial symmetry
is also straightforward.
Let us consider two spherically-symmetric
functions $f(r)$ and $a(r)$ and study the integral
\begin{equation}
h[a](\R) = \int \frac{f(r')
e^{- a(r) |\R-\R'|^2}}{|\R-\R'|} \, d\R' \ .
\end{equation}
Later we can set $a(r) = \omega k_F(r)^2$.
The spherical symmetry allows us to compute
the integral on the $z$-axis alone:
\begin{equation}
\begin{split}
h[a](r) &= \int_0^\infty \int_0^{\pi} 2 \pi r'^2 \sin(\theta) f(r') \\
&\times \frac{e^{-a(r) (r'^2 + r^2 - 2 r' r \cos(\theta))}}
{\sqrt{(r'^2 + r^2 - 2 r' r \cos(\theta))}} \, dr' \, d\theta \ .
\end{split}
\end{equation}
We can use the substitution $a(r) [r'^2 + r^2 - 2 r' r \cos(\theta)] = t^2$
and rewrite the integral over $\theta$ as
\begin{equation}
\begin{split}
\frac{2 \pi f(r') r'}{\sqrt{a(r)} r'}
\left( \int_0^{k_+} e^{-t^2} \, dt - \int_0^{k_-} e^{-t^2}  \, dt \right) \\
= \frac{\pi^{3/2} f(r') r'}{\sqrt{a(r)} r}
\left[ \erf\left(k_+\right) - \erf\left(k_-\right) \right] \ ,
\end{split}
\end{equation}
where we used the symbols $k_+ = \sqrt{a(r)} (r' + r)$ and $k_- = \sqrt{a(r)} |r' - r|$.
Finally, we write the integral as
\begin{equation}
h[a](r) = \frac{\pi^{3/2}}{\sqrt{a(r)} r} \int_0^\infty r' f(r')
\left[ \erf\left(k_+\right) - \erf\left(k_-\right) \right] \, dr' \ .
\end{equation}
We can calculate this expression at $r=0$
(using the Taylor expansion of the error function):
\begin{equation}
h[a](0) = 4 \pi \int_0^\infty r' f(r') e^{-a(0) r'^2} \, dr' \ .
\end{equation}

\end{document}